\documentstyle[aps,prl,graphicx,amssymb]{revtex}
\begin{document}

\draft

\title{Edge of chaos of the classical kicked top map: Sensitivity to initial conditions}

\author{S\'{\i}lvio M. Duarte Queir\'{o}s\thanks{E-mail: sdqueiro@cbpf.br}}
\address{Centro Brasileiro de Pesquisas Fisicas, Rua Xavier Sigaud 150, 22290-180 Rio de Janeiro-RJ, Brazil\\
}
\author{Constantino Tsallis\thanks{E-mail: tsallis@santafe.edu}}
\address{Santa Fe Institute, 1399 Hyde Park Road, Santa Fe, NM, 87501 USA \\
and\\
Centro Brasileiro de Pesquisas Fisicas, Rua Xavier Sigaud 150, 22290-180 Rio de Janeiro-RJ, Brazil}

\date{\today}

\maketitle

\begin{abstract}
We focus on the frontier between the chaotic and regular regions for the
classical version of the quantum kicked top. We show that the sensitivity to the
initial conditions is numerically well characterised by $\xi=e_q^{\lambda_q t} $, where 
$e_{q}^{x}\equiv \left[ 1+\left( 1-q\right) \,x\right]^{\frac{1}{1-q}} \;(e_1^x=e^x)$, and $\lambda_q$ is 
the $q$-generalization of the Lyapunov coefficient, a result that is consistent 
with nonextensive statistical mechanics, based on the entropy 
$S_q=(1- \sum_ip_i^q)/(q-1) \;(S_1 =-\sum_i p_i \ln p_i$). Our analysis
shows that $q$ monotonically increases from zero to unity when  the kicked-top perturbation 
parameter $\alpha $ increases from zero (unperturbed top) to $\alpha_c$, where $\alpha_c \simeq 3.2$. 
The entropic index $q$ remains equal to unity for $\alpha  \ge \alpha_c$, parameter values for 
which the phase space is fully chaotic.
\end{abstract}

\section{Introduction}
Non-linearity is ubiquitously present in nature, e.g., fluid
turbulence\cite{beck}, extinction/survival of species in ecological systems\cite{lotka-volterra}, 
finance\cite{andre-farmer}, the rings of 
Saturn\cite{froyland}, and others. Consistently, the study of low-dimensional
non-linear maps plays a significant role for a better
understanding of complex problems, like the ones just stated. In a classical context, a main characterisation 
of the dynamical state of
a non-linear system consists in the analysis of its sensitivity to initial conditions. 
From this standpoint, the concept of chaos emerged as
tantamount of strong sensitivity to initial conditions\cite{pesin}. In other
words, a system is said to be chaotic if the distance between
two close initial points increases {\it exponentially} in time.
The appropriate theoretical frame to study chaotic and regular behaviour of
non-linear dynamical systems is, since long, well established. It is not so for the region in between, {\it edge of chaos}, 
which has only recently started to be appropriately characterised, by means of the so-called nonextensive
statistical mechanical concepts \cite{gm-ct}.
In this article we study the sensitivity to initial conditions at this  
intermediate region for the classical kicked top map 
and its dependence on the perturbation parameter $\alpha $. 
The sensitivity to initial conditions is defined through  
%
$\xi \left( t\right) \equiv \lim _{\left\Vert \Delta \vec{r}\left( 
0\right) \right\Vert \rightarrow 0}\frac{\left\Vert \Delta \vec{r}%
\left( t\right) \right\Vert }{\left\Vert \Delta \vec{r}\left( 0\right)
\right\Vert }$,  
%
where $\Delta \vec{r}\left( t\right) $ represents the difference, at time $t$, between two
trajectories in phase space. When the system is in a
chaotic state, $\xi $ increases, as stated previously, through an exponential law, i.e.,
\begin{equation}
\xi \left( t\right) \equiv e^{\lambda _{1}\,t}\quad \quad \left( \lambda
_{1}>0\right) ,  \label{sens2}
\end{equation}
where $\lambda _{1}$ is the maximum Lyapunov exponent (the underscript $1$ 
will become transparent right a-head). Equation (\ref{sens2}) can also be regarded 
as the solution of the differential equation, $\frac{d\xi }{dt}=\lambda _{1}\,\xi$ .  
In addition to this, the exponential form has a special relation with the Boltzmann-Gibbs 
entropy $S_{BG} \equiv -\sum_i p_i \ln p_i$. Indeed, the optimization of $S_{BG}$ under 
a fixed mean value of some variable $x$ yields $p(x) \propto e^{- \beta x}$, where $\beta $ 
is the corresponding Lagrange parameter. 
  
Excepting for some pioneering work\cite{mori}, during many years the references to the 
sensitivity at the edge of chaos were restricted to mentioning that it corresponds to 
$\lambda _{1}=0$, with trajectories diverging as a power law of time\cite{pesin}. 
With the emergence of nonextensive statistical mechanics\cite{gm-ct}, a new interest 
on that subject has appeared, and it has been possible to show details, first numerically
\cite{ct-arp-wmz}\cite{ml-ct}\cite{epb-ct-gfja-pmco}\cite{fb-gfja-ct}\cite{gfja-ct} 
and then analytically (for one-dimensional dissipative unimodal maps)\cite{fb-ar}. 
Albeit $\lambda _{1}$ vanishes, it is possible to express the divergence between
trajectories with a form which conveniently generalizes the exponential function, namely 
the $q$-exponential form
\begin{equation}
\xi \left( t\right) 
=\exp _{q_{s}}\left( \lambda
_{q_{s}}\,t\right) \quad \quad \left( \lambda _{q_{s}}>0;\ q_{s}<1\right) ,
\label{q-sens1}
\end{equation}
where $\exp _{q}\,x\equiv \left[ 1+\left( 1-q\right) \,x\right] ^{1/\left(
1-q\right) } \; (\exp _{1}\,x=e^{x})$, and $\lambda _{q_{s}}$ represents the generalised 
Lyapunov coefficient (the subscript $s$ stands for {\it sensitivity})\cite{fb-gfja-ct}.
Equation (\ref{q-sens1}) can be looked as the solution of $\frac{d\xi }{dt}=\lambda 
_{q_{s}}\,\xi ^{q_{s}}$. Analogously to strong chaos (i.e., $\lambda_1>0$), 
if we optimize the entropy \cite{ct}
$S_{q}=\frac{1-\sum\nolimits_{i}p_{i}^{q}}{q-1}   \;\;\;(S_1=S_{BG})$
under the same type of constraint as before, we obtain  $p(x) \propto exp_q(-\beta_q x)$, 
where $\beta_q$ generalizes $\beta$. 
\section{The classical kicked top map}
\label{top}
The classical kicked top corresponds to a map on the unit sphere $x^{2}+y^{2}+z^{2}=1$, corresponding 
to the following application
\begin{eqnarray}
x_{t+1}&=&z_{t}  \nonumber\\ 
y_{t+1}&=&x_{t}\sin \left( \alpha \,z_{t}\right) +y_{t}\cos \left( \alpha \,z_{t}\right) \\ 
z_{t+1}&=&-x\cos \left( \alpha \,z\right) +y_{t}\sin \left( \alpha \,z_{t}\right) \nonumber
\label{mapa}
\end{eqnarray}
where $\alpha $ denotes the \textit{kick strength}. It is straighforward to verify that the
determinant of the Jacobian matrix of (3)
equals one, meaning that this map is  \textit{conservative}. It is therefore quite analogous 
to Hamiltonian conservative systems, the phase space of which consists of a mixing of regular 
(the famous \textit{KAM-tori}\cite{zaslavsky}) 
and chaotic regions characterised, respectively, by a linear ($q_{s}=0$) and exponential 
($q_{s}=1$) time evolution of $\xi $ \cite{fb-gfja-ct}. The region of separation presents 
a $q_{s}$-exponential law for its sensitivity. 
In Fig. 1(a) we exhibit a trajectory where 
the various regions can be seen. In Figs. 1 (b,c,d) we see the time evolution of $\xi $ for 
the three possible stages: regular, edge of chaos and chaotic, respectively. 
\begin{figure}[tbp]
\begin{center}
\includegraphics[width=7cm]{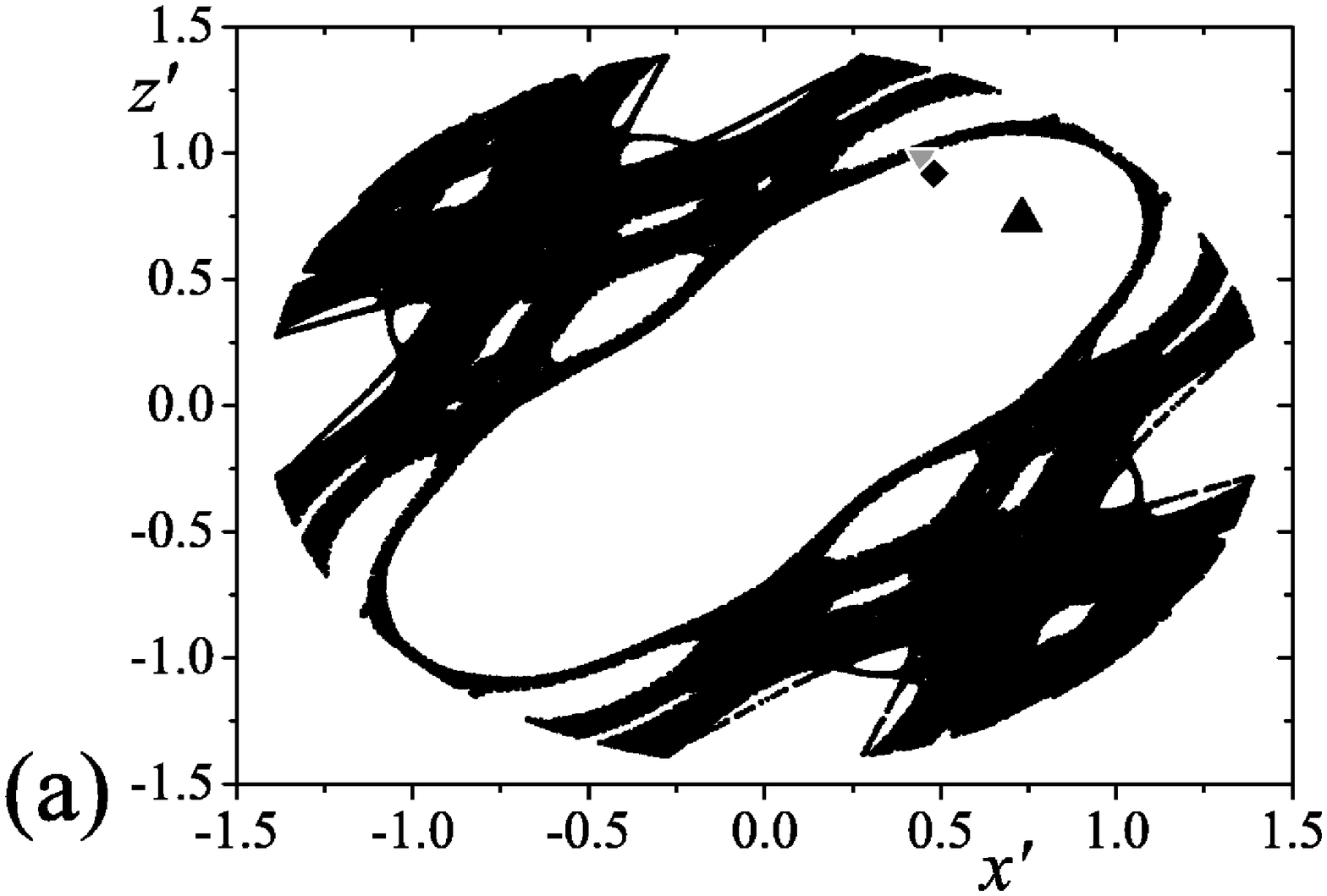}
\includegraphics[width=7cm]{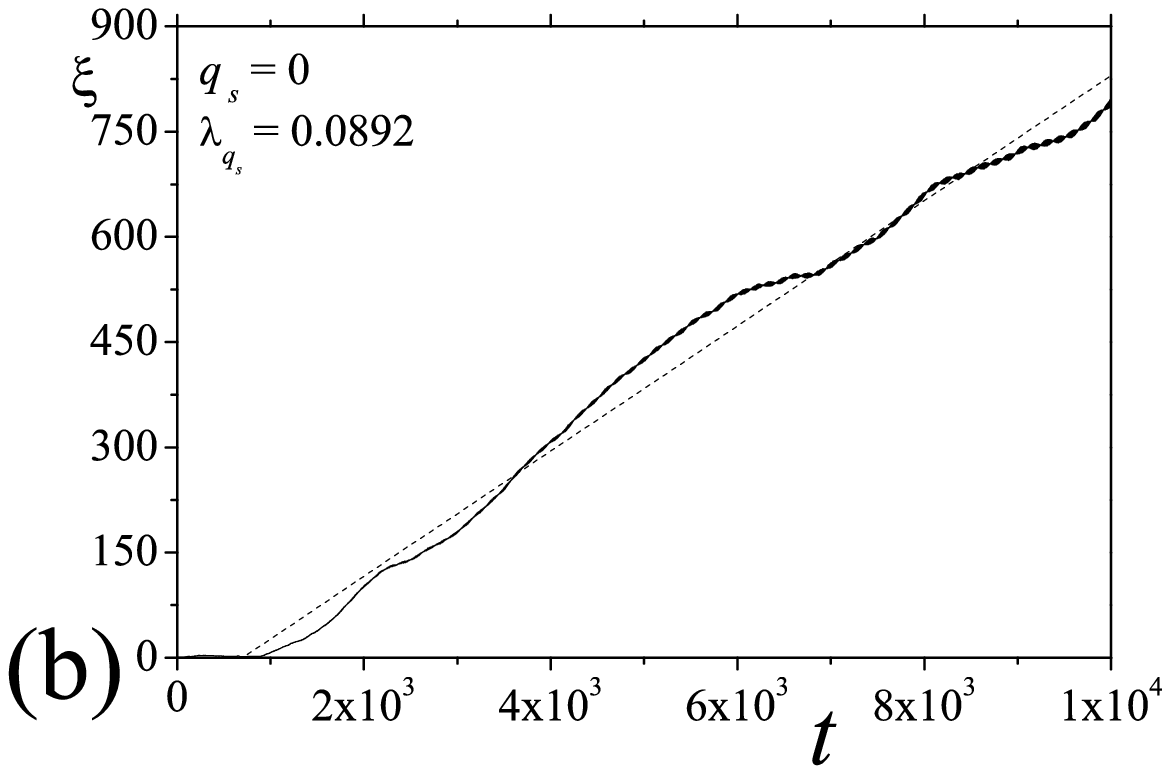}
\includegraphics[width=7cm]{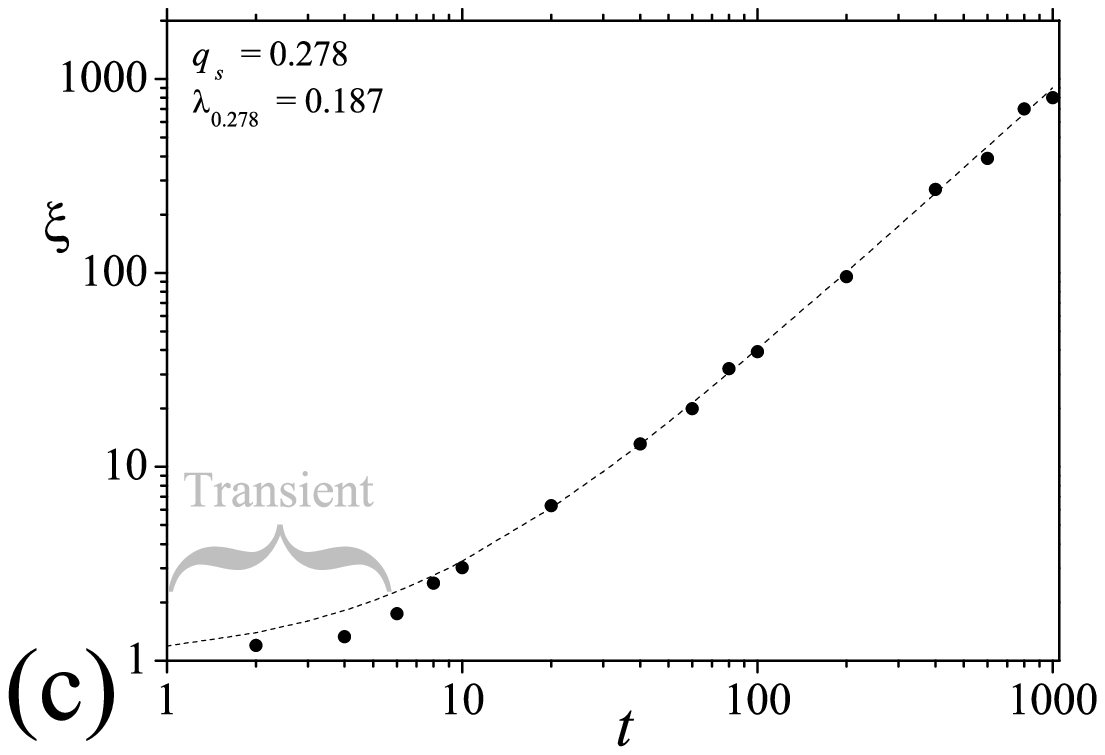}
\includegraphics[width=7cm]{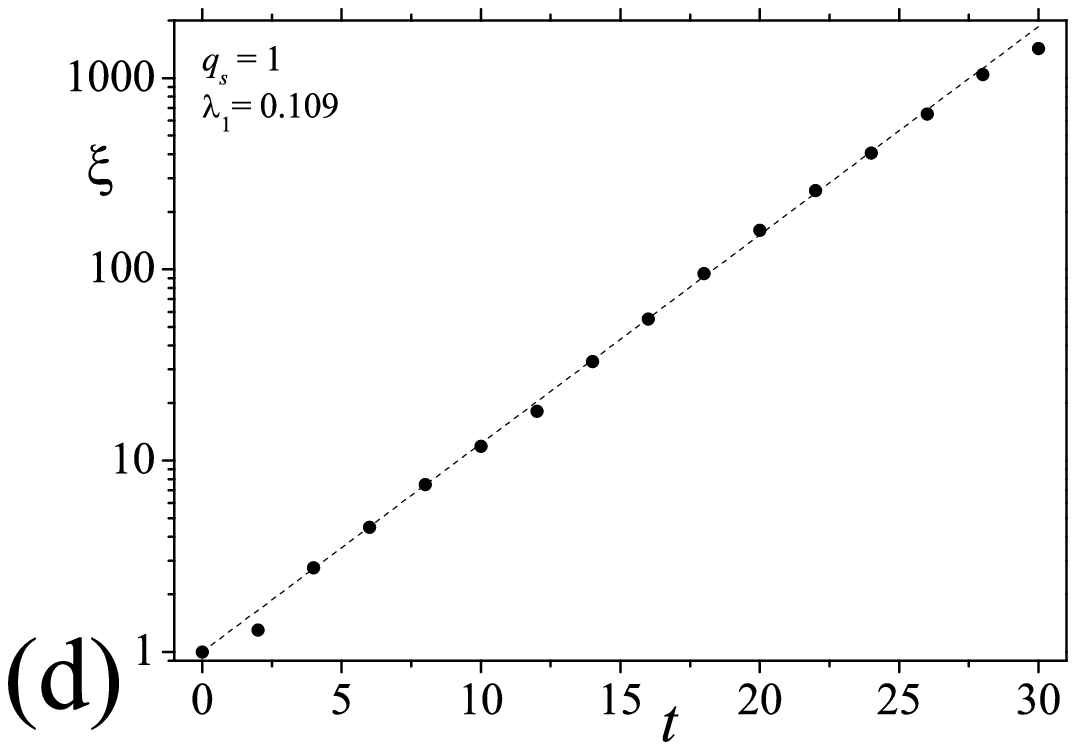}
\end{center}
\caption{(a) Orbit of the $\alpha =2.3$ kicked top, where chaotic and regular regions are visible. 
The spherical phase space is projected onto $x-z$ plane by multiplying the $x$ and $z$ coordinates 
of each point by $R/r$ where $R=\sqrt{2\left(1-\left\vert y\right\vert \right) }$ and 
$r=\sqrt{1-y^{2}}$. (b-d) Time dependence of the sensitivity $\xi$ to initial conditions 
(with $\left\Vert \Delta \vec{r}\left(0\right) \right\Vert = 10^{-10}$) at (b) 
regular region ($\blacktriangle $; {\it linear} evolution), (c) edge of chaos ($\blacklozenge $; 
{\it $q _{s}$-exponential} evolution), and (d) chaotic region ($\blacktriangledown $, {\it exponential} evolution). 
}
\label{fig1}
\end{figure}   
It is worthy mentioning at this point that the quantum version of this map constitutes a paradigmatic 
example of quantum chaos. At its threshold to chaos, it has been verified a nonextensive 
behaviour  (for details see \cite{yw-ct-sl}).

We analysed here the sensitivity to initial conditions on the verge of chaos of (3), for 
several values of the kick strength $\alpha \in [0,4]$ averaged over a set of (typically 50) 
initial conditions for each value of  $\alpha $. More precisely, for fixed $\alpha $, aided 
by its typical orbits, we determined a set of points in the regular-chaos border and then, 
for these points, determined the average value of $\xi $ at time $t$. See typical results 
in Figs. 2 and 3. 
\begin{figure}[tbp]
\begin{center}
\includegraphics[width=7cm]{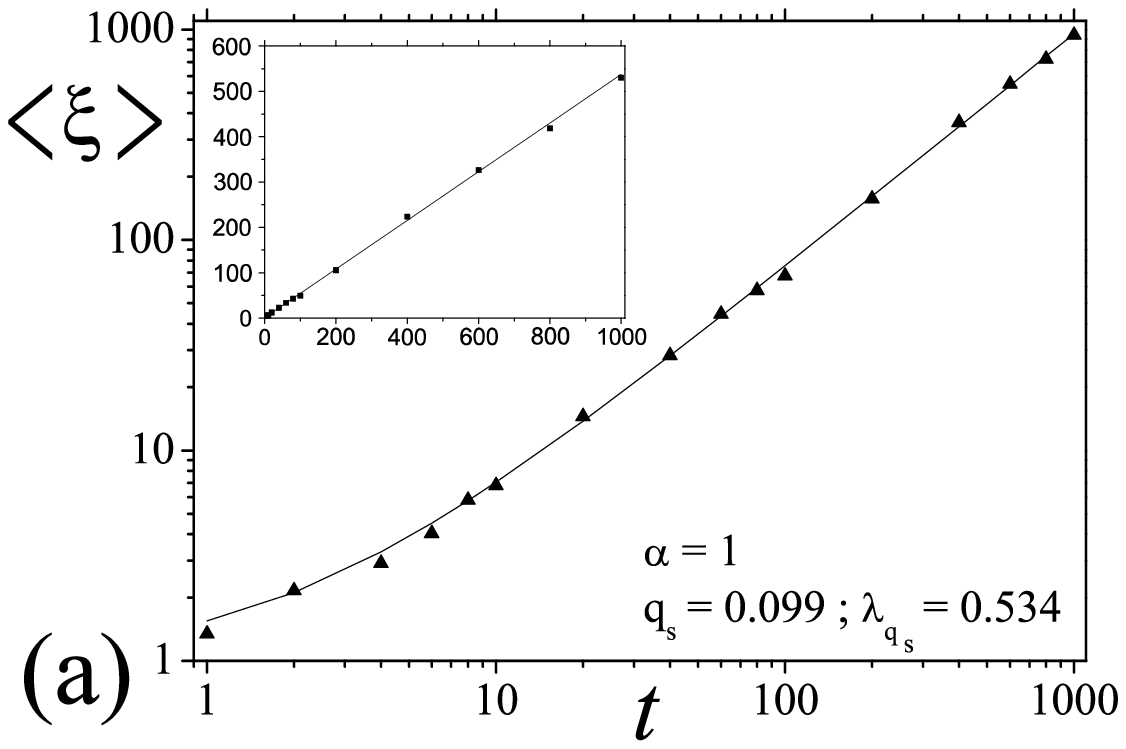}
\includegraphics[width=7cm]{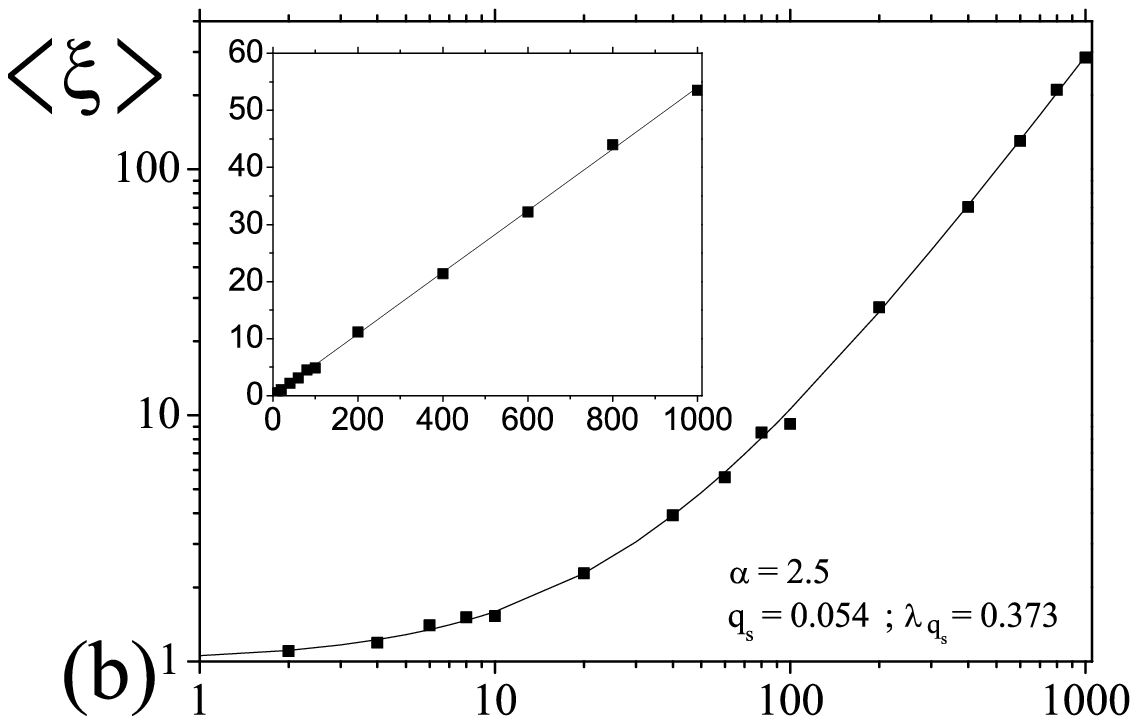}
\includegraphics[width=7cm]{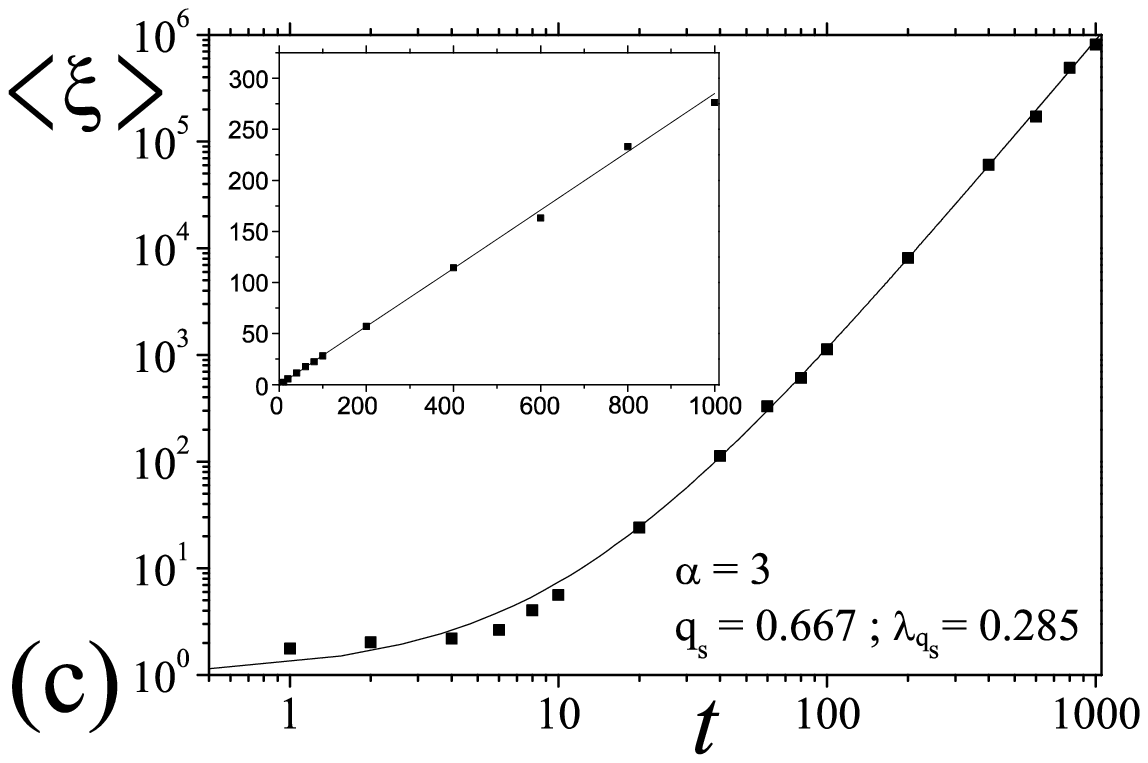}
\includegraphics[width=7cm]{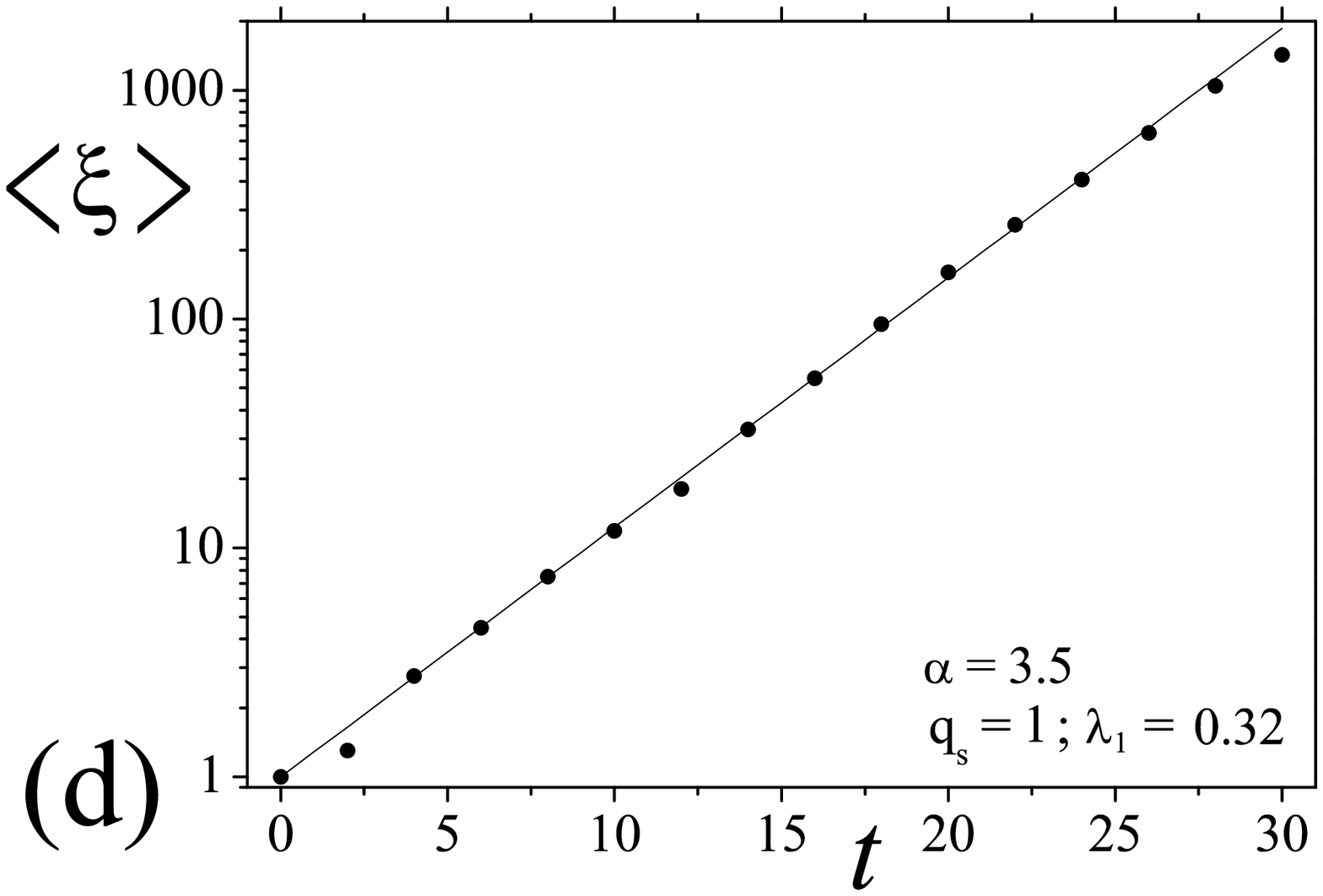}
\end{center}
\caption{Time dependence of the averaged (over close to 50 points of the edge of chaos region) 
sensitivity $\left\langle \xi \right\rangle $ to initial conditions,  for typical values of 
$\alpha $. {\it Insets:} Same data but using a $\ln _{q_{s}}$- ordinate, where 
$\ln _{q}(x) \equiv (x^{1-q}-1)/(1-q)$ ($\ln _{1} = \ln $) . With this  {\it $q$-logarithm} 
ordinate, the slope of the straight line is simply $\lambda _{q_{s}}$.}
\label{fig2}
\end{figure}
We verify that the increase of $\alpha $ induces a gradual approach of 
$q_{s}$ to $1$. This behaviour is in accordance with what was verified\cite{fb-gfja-ct} for 
a non-linear system composed by two simplectically coupled standard maps.
\begin{figure}[tbp]
\begin{center}
\includegraphics[width=15cm]{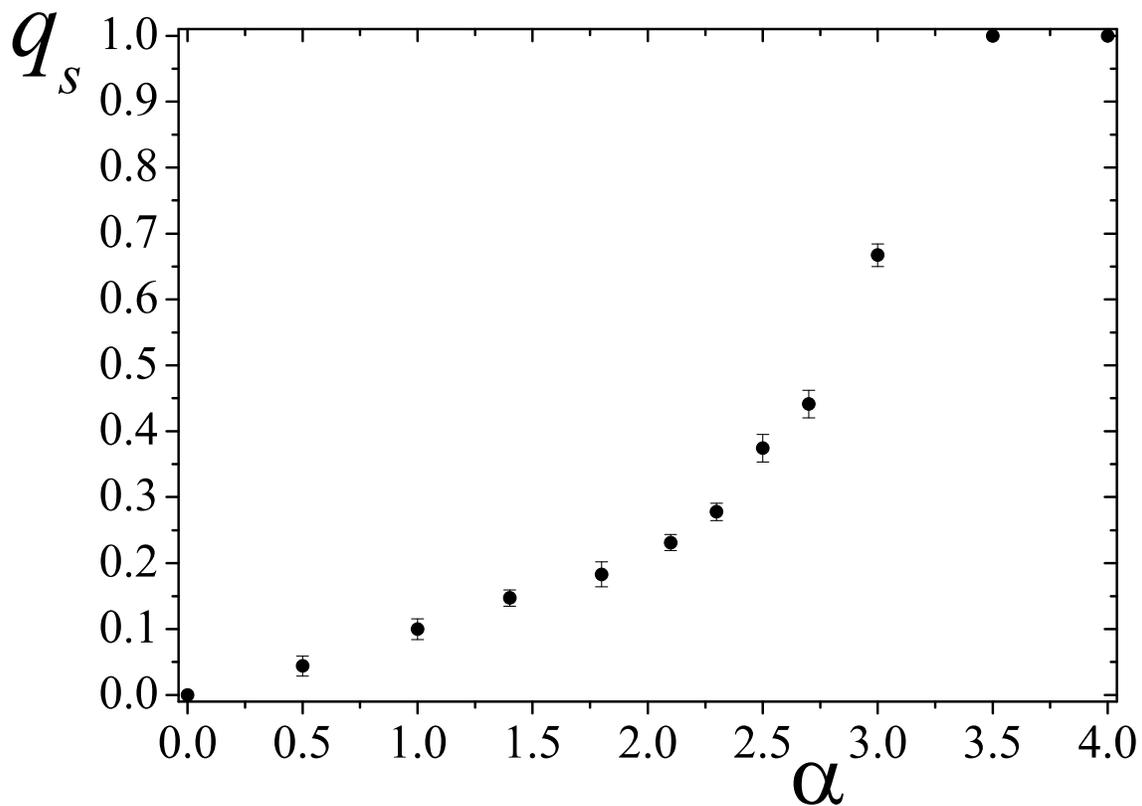}
\end{center}
\caption{The $\alpha$-dependence of $q_{s}$. For some critical value $\alpha_c \simeq 3.2$, $q_{s}$ reaches unity (corresponding  
to fully chaotic phase space), and maintains this value for all $\alpha \ge \alpha_c$.}
\label{fig3}
\end{figure}
\begin{figure}[tbp]
\begin{center}
\includegraphics[width=15cm]{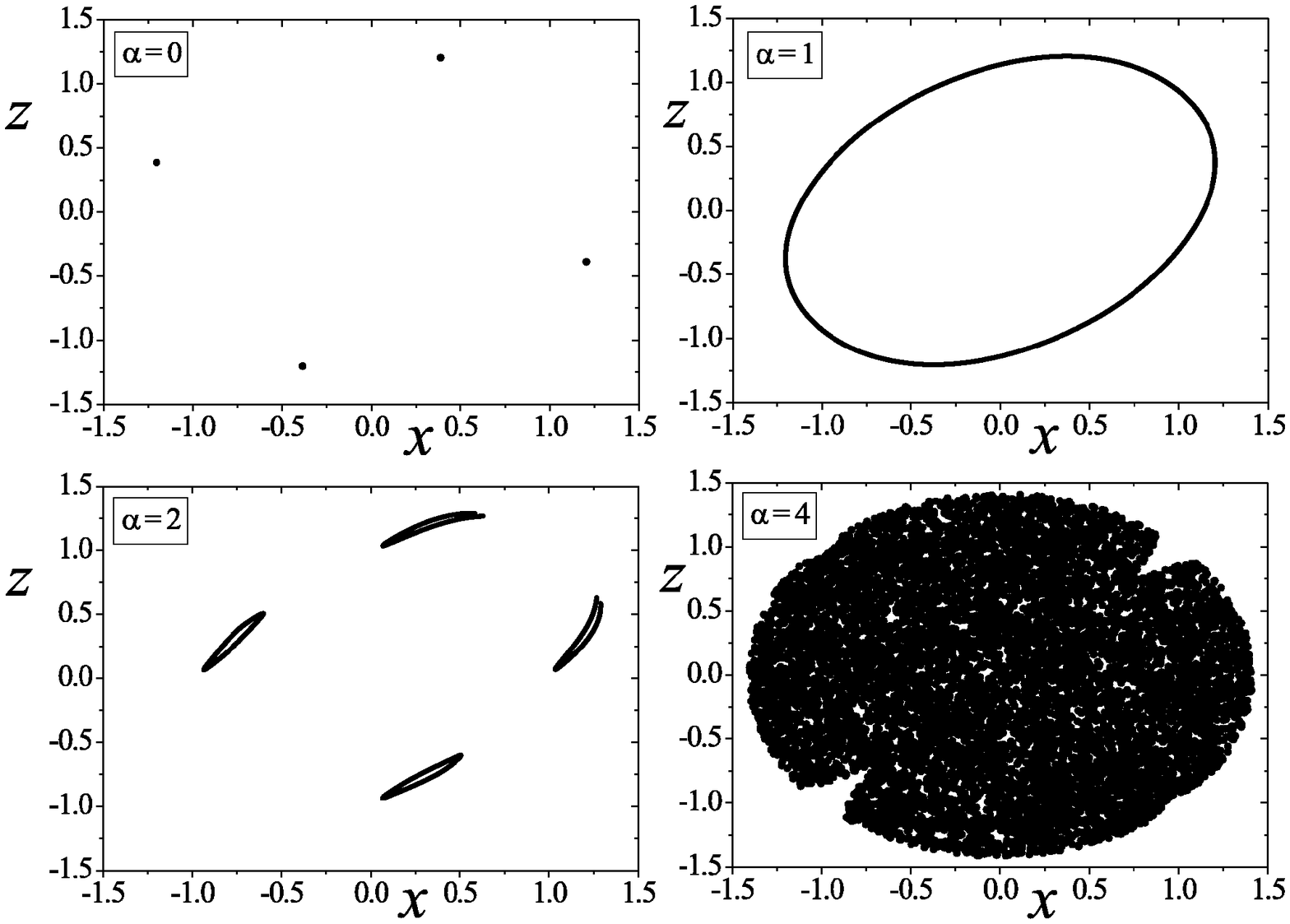}
\end{center}
\caption{Representation of the orbits for typical values of $\alpha $. As $\alpha $
increases we verify the emergence of chaotic regions that,  for $\alpha_c \ge \alpha_c \simeq 3.2$, fulfil the entire phase space.} 
\label{fig4}
\end{figure}

Summarising, we numerically analysed the sensitivity to initial conditions at the edge of 
chaos of the conservative classical kicked top, and found that its time evolution exhibits 
a $q_{s}$-exponential behavior in all cases. For $\alpha =0$, the phase space is composed by a 
regular region, where the sensitivity depends linearly on time, hence $q_{s}=0$. As $\alpha $ 
increases, the top is more perturbed, hence chaotic regions emerge in phase space. Above 
some critical value $\alpha_c \simeq 3.2$, the chaotic region fulfils the entire phase space. 
Consistently, the usual exponential dependence  (i.e., $q_{s}=1$) is recovered. These results 
can be useful to understand, within a nonextensive statistical mechanical framework, the everlasting 
{\it metastable} states that are known to exist in systems composed of many symplectically coupled 
maps\cite{latin}, as well as in isolated many-body long-range-interacting classical Hamiltonians\cite{andrea-fernando}.

\bigskip
We would like to thank F. Baldovin and G.F.J. A\~{n}a\~{n}os for fruitful discussions, as well as FAPERJ, PRONEX 
and CNPq (Brazilian agencies) and FCT/MCES (Portuguese agency) for partial financial support.

\end{document}